\documentclass{IEEEtran}
\usepackage[utf8]{inputenc}
\usepackage[newfloat]{minted}
\usepackage{quantikz}
\usepackage[colorlinks]{hyperref}
\usepackage{svg}
\usepackage{amsmath}
\usepackage{mathtools}

\usepackage{amssymb}
\usepackage{pifont}
\usepackage{graphicx}
\usepackage{algorithmicx} 
\usepackage{algpseudocode}
\usepackage[justification=centering]{caption}
\usepackage{subcaption}
\usepackage{algorithm}
\usepackage{placeins}

\usepackage{array}
\newcolumntype{P}[1]{>{\centering\arraybackslash}p{#1}}
\newcolumntype{M}[1]{>{\centering\arraybackslash}m{#1}}
\newcolumntype{Q}[1]{>{\arraybackslash}m{#1}}
\usepackage{comment}
\usepackage{float}
\usepackage{makecell}
\newcommand{\linebreakand}{%
  \end{@IEEEauthorhalign}
  \hfill\mbox{}\par
  \mbox{}\hfill\begin{@IEEEauthorhalign}
}

\usepackage{tikz}
\usetikzlibrary{shapes.geometric, arrows, decorations.pathmorphing}
\tikzstyle{startstop} = [rectangle, rounded corners,text centered, draw=black, fill=red!30]
\tikzstyle{io} = [trapezium, trapezium left angle=70, trapezium right angle=110, minimum width=3cm, minimum height=1cm, text centered, draw=black, fill=blue!30]
\tikzstyle{process} = [rectangle, text centered, draw=black, fill=orange!30]

\tikzstyle{decision} = [diamond, text centered, draw=black, fill=green!30]
\tikzstyle{arrow} = [thick,->,>=stealth]

\newcommand{\vew}[1]{
	\edef\start{\the\pgfmatrixcurrentrow-\the\pgfmatrixcurrentcolumn}
	\edef\end{\the\numexpr#1+\pgfmatrixcurrentrow\relax-\the\pgfmatrixcurrentcolumn}
	\expandafter\expandafter\expandafter\vewexplicit\expandafter\expandafter\expandafter{\expandafter\start\expandafter}\expandafter{\end}
}
\newcommand{\vewexplicit}[2]{
	\arrow[from=#1,to=#2,arrows,decorate,decoration={snake,amplitude=1pt,segment length=6.5pt}] {}
}
\usepackage{listings}
\usepackage{xcolor}

\definecolor{codegreen}{rgb}{0,0.6,0}
\definecolor{codegray}{rgb}{0.5,0.5,0.5}
\definecolor{codepurple}{rgb}{0.58,0,0.82}
\definecolor{backcolour}{rgb}{0.95,0.95,0.92}

\lstdefinestyle{mystyle}{
    backgroundcolor=\color{backcolour},   
    commentstyle=\color{codegreen},
    keywordstyle=\color{magenta},
    numberstyle=\tiny\color{codegray},
    stringstyle=\color{codepurple},
    basicstyle=\ttfamily\footnotesize,
    breakatwhitespace=false,         
    breaklines=true,                 
    captionpos=b,                    
    keepspaces=true,                 
    numbers=left,                    
    numbersep=5pt,                  
    showspaces=false,                
    showstringspaces=false,
    showtabs=false,                  
    tabsize=2
}

\usepackage{adjustbox}
\usepackage{multirow}
\usepackage{amssymb}
\usepackage{pifont}
\usepackage{siunitx}

\begin{document}
\bstctlcite{IEEEexample:BSTcontrol}

\title{Logical-to-Physical Compilation for Reducing Depth in Distributed Quantum Systems}


\author{%
\IEEEauthorblockN{Folkert de Ronde, Stephan Wong, Sebastian Feld}\\
\IEEEauthorblockA{Quantum \& Computer Engineering\\
Delft University of Technology, Delft, The Netherlands\\
f.w.m.deronde@tudelft.nl, j.s.s.m.wong@tudelft.nl, s.feld@tudelft.nl}
}


\maketitle
\thispagestyle{plain}

\begin{abstract}
Quantum computing is expected to become a foundational technology for solving problems that exceed the capabilities of classical systems. As quantum algorithms and hardware technologies continue to advance, the need for scalable architectures becomes increasingly clear. Distributed quantum computing offers a promising path forward by interconnecting multiple smaller processors into a larger, more powerful system. 
However, distributed quantum computing introduces significant circuit depth overhead, as logical operations are typically decomposed into sequential physical procedures that require entanglement generation. These sequential operations limit the reliability of quantum algorithms in the NISQ era due to noise. In this work, we present a compiler that integrates logical-to-physical decomposition with depth-aware rescheduling to reduce the execution cost of distributed quantum circuits. The compiler identifies sequences of logical CNOT gates that share a control or target qubit, reschedules them into parallel instruction groups, and applies decompositions that allow multiple gates to be executed simultaneously using distributed shared entanglement resources. An algorithm is proposed that ensures parallelism is created when possible while keeping logical equivalence and that circuit depth is never increased. Benchmark results demonstrate that the compiler consistently reduces circuit depth for circuits containing inherently sequential CNOT structures, while leaving already-parallel circuits unchanged. These results highlight the value of combining scheduling and hardware-aware decomposition, and establish the compiler as a practical tool for improving the fidelity of distributed quantum computations.
\end{abstract}

\section{Introduction}\label{sec:intro}

Quantum computing has advanced rapidly in recent years, with new algorithms and hardware platforms pushing the boundaries of what can be achieved. As systems grow in size and complexity \cite{Arute2019QuantumSupremacy, Zhong2020BosonSampling}, it is becoming increasingly clear that future large‑scale quantum computers will not exist as single monolithic processors.
Instead, they will likely be composed of multiple interconnected quantum nodes, which form distributed quantum systems \cite{Monroe2014LargeScale, Nickerson2014DQC, Wehner2018QuantumInternet}. These distributed quantum systems are promising, but it also introduces new challenges that must be addressed at the compiler and system‑software level.
At the same time, we remain in the Noisy Intermediate-Scale Quantum (NISQ) era \cite{Preskill2018NISQ}, where noise and decoherence severely limit the reliability of quantum computations. One of the critical challenges in this context is the depth of quantum circuits: as the number of sequential operations increases, circuits become more vulnerable to noise, reducing the circuit fidelity \cite{Kandala2017HardwareEfficient, Murali2019NoiseAware}. Consequently, reducing circuit depth has become an important part of compiler implementations \cite{Cowtan2019QubitMapping, Murali2019NoiseAware}.

While circuit depth is a challenge across many, if not all, quantum architectures \cite{Nam2020}, it is thus also a problem when considering distributed quantum computing. In distributed systems, the quantum computer is composed of multiple physically separated nodes, where each node has its own qubits, control hardware and the ability to perform local gates. These nodes are in our case connected through photonic links that allow them to share entanglement between electrons and execute non-local operations across the network \cite{Hensen2015LoopholeFree, Pompili2021Multinode} 
. A distributed gate typically requires generating entanglement between the communication qubits of two nodes (electrons in our case) followed by classical communication and conditional operations \cite{Nickerson2014DQC}. When these distributed gates are scheduled sequentially, they can dramatically increase circuit depth, amplifying the effects of noise and limiting scalability.

\begin{figure}
    \centering
    \includegraphics[width=0.9\linewidth]{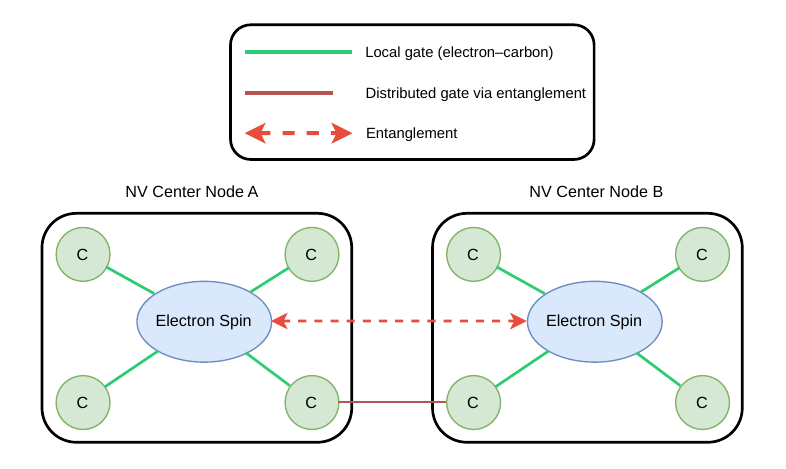}
    \caption{Two NV‑center nodes in a distributed quantum computer. Each node contains an electron spin used as a communication qubit and four C‑13 nuclear spins used as memory qubits. Entanglement is generated between the electron spins of different nodes via photonic links, enabling distributed quantum gates between the carbon qubits.}
    \label{fig:NV_center_show}
\end{figure}

This challenge becomes even more critical when executing logical quantum algorithms on physical hardware. Logical operations in an algorithm must be decomposed into hardware‑native physical instructions, a process known as logical‑to‑physical decomposition \cite{Chamberland2022Topological}. In distributed architectures, this decomposition often expands a single logical CNOT into a multi‑step protocol involving entanglement generation, measurement, and conditional corrections across nodes \cite{Nickerson2014DQC, Monroe2014LargeScale}. An example of a distributed system is shown in Figure \ref{fig:NV_center_show}. As a result, even simple logical structures can translate into long physical sequences, making depth reduction essential for maintaining fidelity and enabling scalable distributed quantum computation.

Although logical-to-physical decomposition can turn even simple logical operations into long sequences of physical gates, these sequences are not always required to remain sequential once the operations are decomposed into their physical implementations. Some logical operations that appear strictly sequential at the algorithmic level are capable of being executed in parallel on the underlying hardware. Logical CNOT gates that share a control qubit or share a target qubit can often be executed in parallel using a single entanglement resource. 
In these situations, the logical circuit can give the impression that certain operations must occur one after another, even though the underlying hardware is capable of executing several of them concurrently. The physical system can support this parallelism, provided the compiler can identify and manage this parallelism effectively. Realizing this potential requires a compiler that can identify these structural opportunities, reorder gates without violating logical correctness, and apply joint decompositions that exploit shared entanglement resources to reduce physical depth. Existing compilers lack these capabilities \cite{Sivarajah2020Tket, Murali2019NoiseAware}, leaving parallelism potentials unused.

In this work, we present a compiler that reduces circuit depth by uncovering parallelism in distributed quantum operations. Logical gates that would normally be treated as strictly sequential can, through our decomposition strategy, be restructured to be executed in parallel. The compiler automatically identifies when such structures occur and reorders instructions to exploit this parallelism, allowing distributed gates to run in parallel rather than in series. The compiler then applies CNOT decompositions that use fewer physical operations and achieve meaningful depth reduction without requiring any additional entanglement resources. This approach directly addresses one of the central challenges of NISQ computing: mitigating noise and improving fidelity through shallower circuits.

The research question addressed in this paper is:
\textbf{What role can compiler-based decomposition play in detecting and exploiting parallelism in seemingly sequential distributed quantum circuits?}

The contributions of this work are as follows:
\begin{enumerate}
    \item 
    We introduce a method by which logical operations that appear sequential can be decomposed into parallel execution, revealing parallelism in distributed quantum circuits.

    \item 
    We design and implement a compiler capable of identifying these parallelizable structures and scheduling them accordingly, ensuring that distributed gates are executed in parallel if possible.

    \item 
    By decomposing and rescheduling distributed operations, particularly CNOT gates, our approach significantly reduces circuit depth, thereby improving fidelity in NISQ systems.
\end{enumerate}

The remainder of the paper is structured as follows. In Section \ref{sec:related_work}, we compare our approach with existing compiler optimizations and highlight the gap in strategies for distributed gate execution. 
In Section \ref{sec:Compiler_design}, we present the theoretical foundations of our method, characterizing the structural patterns that permit parallel execution of distributed CNOT gates and establishing the correctness conditions for the joint decompositions used by our compiler.
Section \ref{sec:compiler_implementation} presents the design of our compiler, including the algorithms used for instruction reordering. Section \ref{Sec:results_discussion} reports our results, demonstrating the reduction in circuit depth. Finally, Section \ref{sec:conclusion} summarizes our findings and outlines directions for future research.

\section{Related Work}\label{sec:related_work}
Research on quantum compilers has focused on a variety of optimization strategies aimed at improving circuit fidelity and execution efficiency. Early efforts concentrated on generic transpilation frameworks that translate quantum circuits into hardware-compatible instructions, often emphasizing gate count reduction and qubit routing \cite{Amy2013, Smith2019Qiskit, Sivarajah2020Tket, Maslov2018, Nash2020}.
Hardware-aware compilers have extended these ideas by exploiting system-specific properties, such as connectivity constraints or native gate sets, to achieve more efficient execution \cite{Murali2019NoiseAware, Li2020QubitMapping, Cowtan2019QubitMapping, Zulehner2018, Childs2019},

In parallel, distributed quantum computing has emerged as a promising approach for scaling quantum systems beyond the limits of single devices. 
Several works have explored protocols for entanglement generation and distributed gate execution across multiple nodes \cite{Nickerson2014DQC, Monroe2014LargeScale, Pompili2021Multinode, Cirac1999, VanMeter2016, Caleffi2018}. While these studies highlight the feasibility of distributed architectures, they often assume sequential scheduling of distributed operations, which can lead to excessive circuit depth and reduced fidelity. 

Recent compiler research has begun to address depth reduction more directly, with techniques such as gate reordering, commutativity-based optimizations, and parallel scheduling \cite{Zhou2020Commutation, Chakrabarti2021SABRE, Tan2023DepthOptimized, Nam2018, Younis2022, Bertoni2023}. However, existing approaches typically focus on local gates within a single device, leaving distributed operations underexplored. To the best of our knowledge, no prior work has systematically investigated compiler strategies for identifying and decomposing distributed CNOT gates that seem to be sequential into parallel execution.

Our work expands on these foundations by introducing a compiler that specifically targets distributed CNOT gates that seem to be sequential, automatically detecting when sequential structures can be decomposed into parallel execution. This fills a gap between general-purpose depth reduction techniques and the unique requirements of distributed quantum architectures.

\section{Theoretical foundation} \label{sec:Compiler_design}
\begin{figure*}
    \centering
    \includegraphics[width=0.8\linewidth]{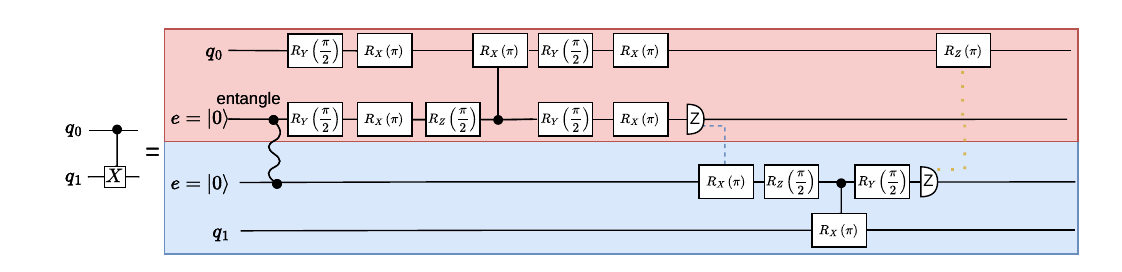}
    \caption{Standard implementation of a distributed CNOT gate. The logical CNOT is decomposed into physical operations involving entanglement generation, classical communication, and physical qubit operations. This construction serves as the baseline against which parallel decompositions are compared.}
    \label{fig:dist_CNOT}
\end{figure*}

Distributed quantum computing enables logical operations to be executed across physically separated nodes, but this flexibility introduces significant overhead. Non-local gates require entanglement generation and sometimes they require classical communication. In standard logical to physical decompositions, each distributed CNOT is treated as an isolated, sequential operation with its own entangled photon pair. As illustrated in Fig. \ref{fig:dist_CNOT}, this leads to physical circuits with a high circuit depth and high entanglement consumption when multiple CNOTs appear in sequence.

In this section, we explain why many of these sequential logical CNOT gates are not inherently sequential at the physical level. When multiple CNOTs share structural relationships, such as a common control or a common target qubit, they can be decomposed into a form that allows them to be executed in parallel using shared entanglement resources. This observation forms the theoretical foundation for the compiler presented in this paper.

\begin{figure*}
    \centering
    \includegraphics[width=1\linewidth]{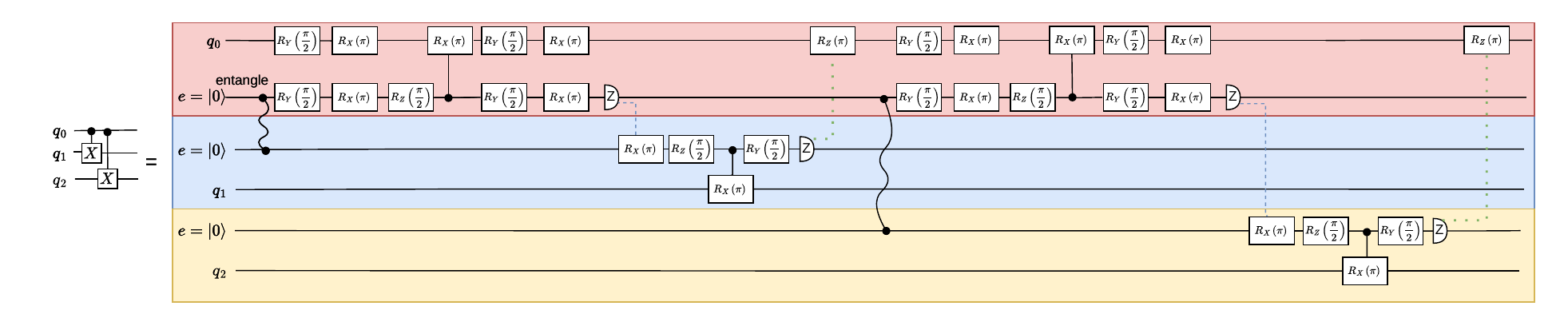}
    \caption{Standard implementation of two distributed CNOT gates sharing a common control qubit. Each logical CNOT is decomposed into a sequence of physical operations involving entanglement generation, classical communication, and local qubit operations. Because both gates use the same control qubit, their physical decompositions must be executed sequentially, illustrating the baseline case against which parallel decompositions are compared.}
    \label{fig:dist_CNOT_2}
\end{figure*}

The conventional distributed CNOT protocol shown in Fig. \ref{fig:dist_CNOT} illustrates how a single logical CNOT is implemented using entanglement generation, classical communication, and local operations. When multiple distributed CNOT gates appear in a circuit, each of these require their own entangled-photon pair to enable the distributed interaction. As a result, executing several such gates in sequence leads to repeated entanglement generation and a long chain of dependent physical operations, as shown in Fig. \ref{fig:dist_CNOT_2}. This causes both circuit depth and entanglement consumption to scale poorly, making the standard construction an inefficient baseline for distributed computation.




To address this limitation, it is desirable to execute multiple distributed CNOT gates simultaneously. For two specific CNOT configurations, this can be achieved using a shared entangled photon pair.

\begin{figure}
    \centering
    \includegraphics[width=1\linewidth]{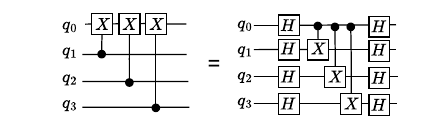}
    \caption{Alternative decomposition enabling parallel execution of distributed CNOT gates that share a target qubit.}
    \label{fig:dist_CNOT_multiple_nodes_shared_target}
\end{figure}

The second case arises when multiple distributed CNOT gates share the same target qubit. In this scenario, the shared‑control decomposition cannot be applied directly, as the structure of the protocol does not generalize to a shared target across multiple nodes. To address this limitation, we introduce an alternative construction, shown in Fig.~\ref{fig:dist_CNOT_multiple_nodes_shared_target}, which enables parallel execution for circuits with a shared target qubit as well.

\begin{figure*}
    \centering
    \includegraphics[width=0.8\linewidth]{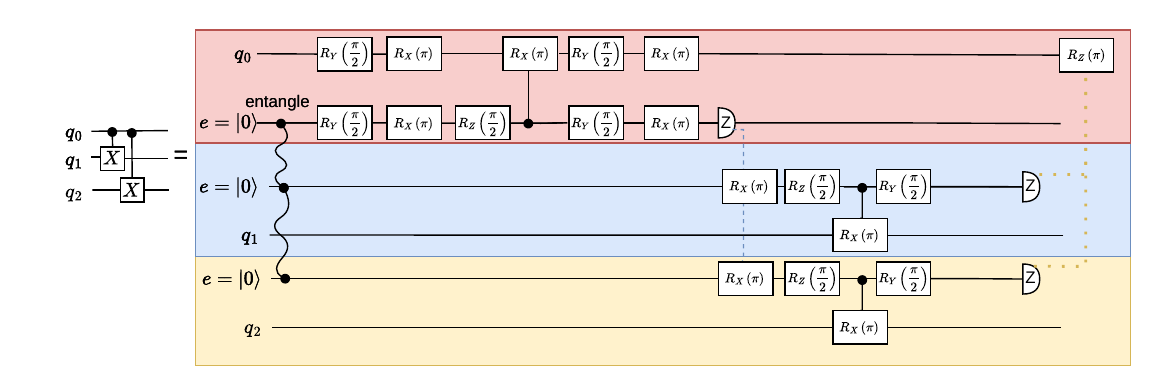}
    \caption{Parallel decomposition of n number of distributed CNOT gates that share a control qubit across n number of nodes. A shared multipartite entangled state enables all CNOTs to be executed simultaneously, reducing circuit depth without increasing entanglement requirements.}
    \label{fig:dist_CNOT_multiple_nodes_shared_control}
\end{figure*}
\begin{figure}
    \centering
    \includegraphics[width=0.9\linewidth]{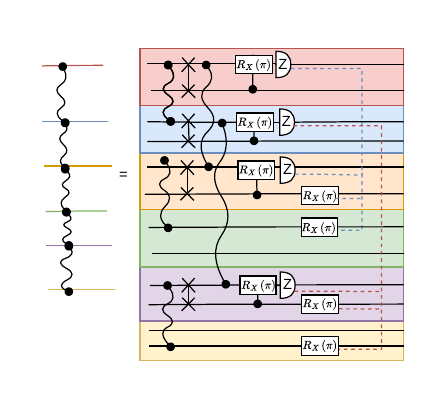}
    \caption{Entanglement creation protocol used to distribute a multipartite entangled state among all participating nodes. This shared resource enables the parallel execution of distributed CNOT gates described in Figs. \ref{fig:dist_CNOT_multiple_nodes_shared_control} and \ref{fig:dist_CNOT_multiple_nodes_shared_target}.}
    \label{fig:entanglement_creation}
\end{figure}

To carry out the parallel decompositions, both for circuits with shared control qubits and those with shared target qubits, it is necessary for all participating nodes to share entanglement. The constructions shown in Figs.~\ref{fig:dist_CNOT_multiple_nodes_shared_control} and \ref{fig:dist_CNOT_multiple_nodes_shared_target} assume the availability of such multipartite entanglement as a common resource. This shared entanglement is established using the entanglement-creation procedure shown in Fig.~\ref{fig:entanglement_creation}, which prepares the required multipartite state that enables the parallel execution of the distributed CNOT operations.


The constructions in Figs.~\ref{fig:dist_CNOT_multiple_nodes_shared_control} and \ref{fig:dist_CNOT_multiple_nodes_shared_target} rely on a shared multipartite entangled state to enable simultaneous execution of multiple distributed CNOT operations.
When combining the creation of entangled pairs with the parallel decompositions of the logical gates, the resulting strategy achieves a more favorable depth scaling growing by only one instruction per additional gate rather than the 19-instruction increase of the conventional approach.

In our method, the overall depth is fixed at 42 instructions. By contrast, the standard sequential protocol adds 19 instructions per distributed CNOT. Although our approach still scales with the number of gates, the growth is dramatically slower, making the parallel decomposition advantageous whenever three or more distributed CNOT gates appear in sequence.

\begin{figure}
    \centering
    \includegraphics[width=0.8\linewidth]{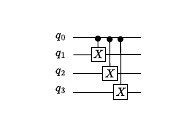}
    \caption{Example of a sequence of CNOT gates that all share the same control qubit. Although these gates appear sequential in the logical circuit, their shared-control structure allows them to be decomposed into a parallel physical implementation.}
    \label{fig:dist_CNOT_multiple_control_logical}
\end{figure}

\section{Compiler implementation} \label{sec:compiler_implementation}


The goal of the compiler is to translate logical quantum circuits into quantum circuits that consist only of physical gates that minimize circuit depth, particularly in the presence of distributed CNOT gates. Achieving this requires more than simply decomposing logical gates: the compiler must also reschedule operations to expose parallelism and group CNOT gates that can be jointly decomposed using the parallel constructions described in Section \ref{sec:Compiler_design}.

The optimization process begins by constructing parallel groups of instructions we call ``buckets'' using the standard procedure for scheduling quantum circuits: each instruction is placed into the earliest bucket in which it can be executed without violating hardware constraints. At this stage, distributed CNOT gates that share a control or target qubit are still treated as sequential operations and therefore end up in separate buckets. This initial bucketization step provides a faithful baseline execution order, ensuring that any subsequent optimization can only maintain or improve upon the naive decomposition.

Once the circuit has been organized into these buckets, the compiler scans the sequence to locate consecutive buckets containing distributed CNOT gates that share either a common control qubit or a common target qubit, since these are precisely the patterns that can be replaced by a parallel decomposition.
The operations within these buckets cannot be rearranged arbitrarily: moving individual operations between buckets risks destroying parallelism and may inadvertently increase circuit depth. If only a subset of the CNOT gates in a bucket can be shifted to a different bucket, the original bucket itself still remains in place and continues to contribute a full depth layer to the circuit. In such cases, no depth reduction is achieved. To avoid this, the compiler treats each bucket of CNOT gates as an indivisible unit, ensuring that only transformations that preserve the full parallel structure of a bucket are considered. Single-qubit operations are excluded from this restriction. Because the decomposition of a distributed CNOT gate has far more circuit depth and is more resource-intensive than that of a single‑qubit gate, allowing single-qubit operations to remain sequential while parallelizing multiple CNOT gates still yields a net reduction in overall circuit depth.

\begin{algorithm}
\caption{Distributed CNOT Parallelization Compiler}
\label{alg:compiler}
\begin{algorithmic}[1]

\Require Quantum circuit $C$
\Ensure Optimized circuit with parallelized distributed CNOT gates

\State Partition all instructions in $C$ into parallel buckets $B = \{B_1, B_2, \dots, B_n\}$
\State Let each CNOT gate be represented as $g=(c,t)$, where $c$ is the control qubit and $t$ is the target qubit

\Comment{Forward sweep}

\For{all buckets $B_i \in B$}
    \For{all CNOT gates $g=(c,t) \in B_i$}
    
        \State movable\_set $\leftarrow \emptyset$
        \State DestBuckets $\leftarrow \emptyset$
        \State $Q \leftarrow$ set of qubits used in Buckets\_seen

        \For{each bucket $B_j \in$ Buckets\_seen}

            \State movable $\leftarrow$ TRUE
        
            \If{$c \in Q$ or $t \in Q$ and dependency caused by single-qubit gate}
                \State movable $\leftarrow$ FALSE

            \ElsIf{$c \in Q$ and $t \in Q$}
                \State movable $\leftarrow$ FALSE
            \EndIf

            \If{movable}
                \If{$g$ does not share a control or target qubit with any CNOT in $B_j$}
                    \State Add $g$ to movable\_set
                    \State Add $B_j$ to DestBuckets
                    \State \textbf{break}
                \Else
                    \State \textbf{continue}
                \EndIf
            \EndIf

        \EndFor
    \EndFor
\EndFor

\If{all CNOT gates in $B_i$ are in movable\_set}
    \State Move all CNOT gates from $B_i$ to their corresponding bucket in DestBuckets
\EndIf

\Comment{Backward sweep}
\State Do same as forward sweep in reverse direction

\Comment{Cleanup and re-bucketing}
\State Remove all empty buckets from $B$
\State Recompute parallel buckets to maximize parallelism

\Comment{Gate decomposition}
\For{all buckets $B_i \in B$}
    \State Identify CNOT groups with shared control or shared target qubits
    \State Sort CNOT gates by qubit indices with shared qubits prioritized
    \While{$B_i$ not empty}
        \State Extract maximal group of CNOT gates sharing a control or target qubit
        \State Decompose group using corresponding distributed CNOT decomposition
    \EndWhile
\EndFor

\State \textbf{return} optimized circuit

\end{algorithmic}
\end{algorithm}

With the bucket structure in place, the compiler proceeds through four passes to identify and restructure sequences of distributed CNOT gates. In the first pass, the compiler walks through the logical circuit and assigns each instruction to the earliest bucket in which it can be executed without violating hardware constraints. 

In the second pass, the compiler iterates over the buckets in order and examines the CNOT gates they contain. For each bucket, every CNOT gate is checked for a shared control or shared target qubit with the CNOT gates in the immediately preceding bucket. Whenever this condition is met, the gate is marked as a candidate for relocation. If all CNOT gates in the current bucket satisfy this criterion, the entire bucket is merged into the preceding one. 


After the forward sweep completes, the third pass performs a reverse sweep using the same logic. This backward pass allows the compiler to detect additional opportunities for merging that were not visible when scanning in the forward direction. As before, only complete buckets are moved, preserving the parallel structure established during initial scheduling.



Finally, the fourth pass recomputes the parallel buckets from the updated instruction sequence. This step accounts for all bucket merges performed in the previous passes and may reveal new opportunities for parallel execution. Any remaining empty buckets are removed, yielding a quantum circuit rescheduled for depth-reduced logical to physical decomposition that is guaranteed not to exceed the depth of the naive decomposition.


After these four passes, the compiler has identified all regions where distributed CNOT gates could, in principle, be merged into a single parallel decomposition. A key design requirement is that applying such a transformation must never increase circuit depth relative to the naive decomposition, in which each distributed CNOT expands into a fully sequential sequence of physical operations. To enforce this guarantee, the compiler evaluates each candidate region before performing any decomposition. The depth impact can be determined directly from the number of logical CNOT gates involved: whenever more than two distributed CNOTs appear in a sequence of consecutive buckets, the parallel construction is guaranteed to reduce depth compared to the naive expansion. This criterion allows the compiler to decide, for every identified region, whether a parallel decomposition is beneficial or whether the naive decomposition should be retained, ensuring that the final output circuit is always depth‑optimal with respect to the available decomposition strategies.



Once the compiler has determined which regions of the circuit benefit from a parallel decomposition, the resulting buckets are expanded into their corresponding physical implementations. Single-qubit gates are decomposed using standard, low-depth procedures and are not the primary focus of this work. The main optimization effort instead targets the decomposition of distributed CNOT gates.

During the earlier compiler passes, buckets may contain CNOT gates that fall into any of the three parallelizable patterns: gates that are fully independent, gates that share a control qubit, or gates that share a target qubit. These patterns can safely coexist within the same bucket because the passes only ensure that all gates in the bucket can be executed in parallel. The decomposition stage must therefore distinguish between these cases and apply the correct physical implementation.

Distributed CNOTs can be decomposed in three distinct ways:
\begin{itemize}
\item fully in parallel,
\item using a shared control qubit, or
\item using a shared target qubit.
\end{itemize}

The specific decompositions are presented in Section \ref{sec:Compiler_design}. To apply them correctly, the compiler must first determine which type of parallel pattern is present within each bucket. The decomposition stage must now identify which form actually applies. To do this, the compiler sorts the CNOT gates in each bucket by their qubit indices, which naturally groups together gates that share a control qubit or share a target qubit. It then processes the sorted list sequentially, collecting gates into contiguous groups that correspond to one of the supported parallel patterns. Once the groups are established and their type is known, the compiler decomposes each group as a single unit. This guarantees that all CNOT gates in the group are expanded together using the appropriate parallel construction, enabling the compiler to realize the full depth reduction exposed by the earlier passes.

Taken together, these stages form a complete pipeline for translating a logical circuit into a depth‑optimized physical implementation. The compiler first exposes all available parallelism through careful bucketization, then restructures the circuit through forward and backward passes to identify regions where distributed CNOT gates can be executed jointly. Finally, it classifies and decomposes each bucket according to the appropriate parallel pattern, ensuring that every optimization opportunity uncovered earlier is fully realized. The resulting physical circuit respects all hardware constraints while achieving the minimum depth achievable by the available decomposition and compiler strategies.

\section{Evaluation and discussion} \label{Sec:results_discussion}
To assess the effectiveness of the theory and the compiler, we evaluated it on a suite of benchmarks. 
The primary goal of the evaluation is to determine whether the compiler's transformations reduce the circuit depth after logical-to-physical decomposition while remaining logically equivalent to the original circuit.

We evaluated the compiler on 15 real benchmark circuits extracted from MQTBench \cite{quetschlich2023mqtbench}, these were the benchmarks that could be extracted for maximum qubit number, and 1000 synthetic circuits generated using ketGPT \cite{ketGPT} containing varying numbers of distributed CNOT gates, with a minimum of 129 and a maximum of 35886 CNOT gates, arranged in different structural patterns. These benchmarks were used to test whether the compiler can detect opportunities for parallel execution when distributed CNOT operations share either a control qubit or a target qubit.

\begin{figure}
    \centering
    \includegraphics[width=1\linewidth]{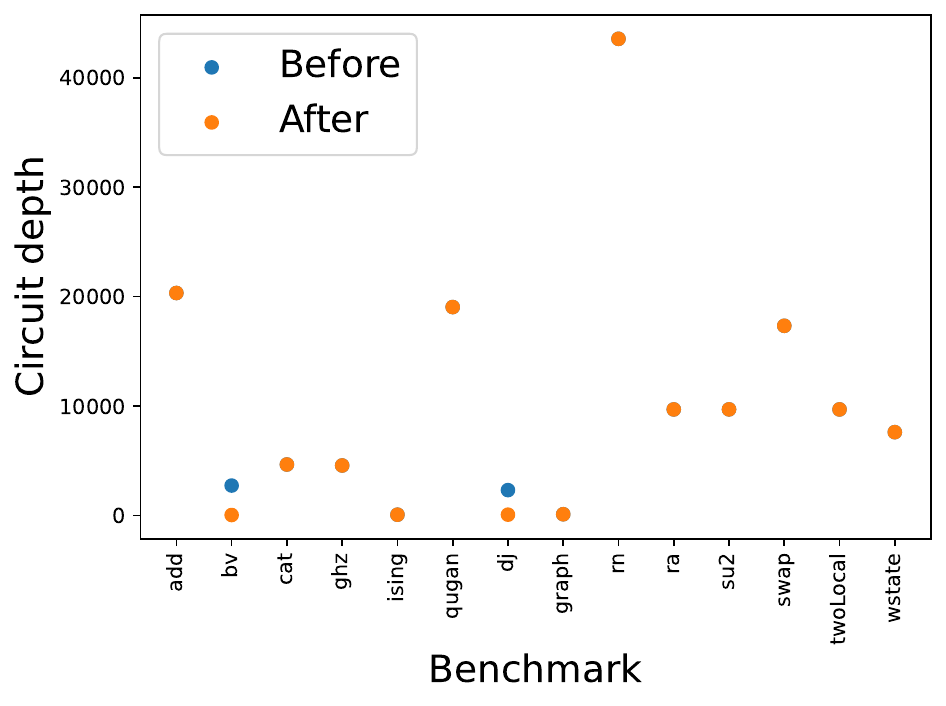}
    \caption{Benchmark results comparing naive sequential decomposition (blue) with the compiler's parallelized output (orange). The compiler consistently reduces circuit depth when sequential distributed CNOT structures are present, while never increasing depth in any benchmark.}
    \label{fig:depth_values}
\end{figure}

Across all benchmarks, the compiler successfully detected sequences of distributed CNOT gates that shared either a control or a target qubit. The results (presented in Fig. \ref{fig:depth_values}) confirm that the compiler’s multi-pass bucket merging strategy exposes parallelism that is not visible in the original logical circuit. The compiler exploits commutation-based parallelism in a conservative way, i.e., it apply only transformations that preserve the guarantee that circuit depth will not increase. For example, even if the control qubits of CNOT gates may commute with rz gates, such commutations do not always yield a depth reduction. In all tested cases, the compiler's transformations preserved circuit semantics and produced valid physical implementations.


The primary performance metric considered in this evaluation and work is the resulting circuit depth after decomposition into the physical gate set. Figure~\ref{fig:depth_values} compares the depth obtained using a naive sequential decomposition (marked blue) with the depth produced by the compiler's optimized parallel decomposition (marked orange).
As expected, the compiler reduces circuit depth whenever multiple distributed CNOT gates appear with either shared target or control qubits, as can be seen for the Deutsch-Jozsa (DJ) and Bernstein–Vazirani (BV) circuit. If the logical circuit already exhibits maximal parallelism, the compiler correctly refrains from applying unnecessary transformations. In such cases, the circuit depth remains unchanged (represented by just an orange dot in the figure). Notably, none of the evaluated benchmarks showed an increase in circuit depth, confirming that the compiler's constraints prevent transformations that would negatively impact the circuit depth.

\begin{figure}
    \centering
    \includegraphics[width=1\linewidth]{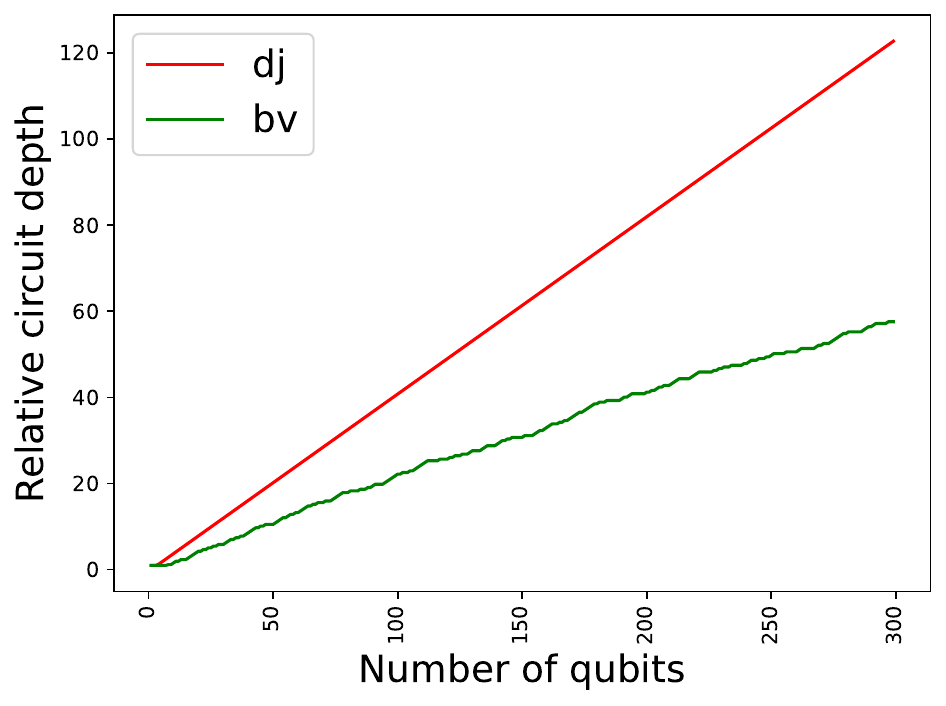}
    \caption{Relative circuit depth improvements achieved by the compiler for the Deutsch-Jozsa (DJ) and Bernstein-Vazirani (BV) circuits as the number of qubits increases. The results show that the relative depth reduction grows with circuit size, demonstrating that the compiler's optimization strategy scales effectively for larger circuits.}
    \label{fig:mqt_bench_absolute_relative}
\end{figure}

To study how the benefits of the compiler scale with circuit size, we selected the benchmark circuits from Figure \ref{fig:depth_values} that showed improvements after optimization (DJ and BV). These circuits were then evaluated for increasing numbers of qubits.
Figure \ref{fig:mqt_bench_absolute_relative} shows the relative circuit depth reduction achieved by the compiler as the number of qubits increases (and therefore also the circuit size) for the DJ and BV circuits. The results demonstrate that the benefits of the compiler grows as the number of qubits grows. In both benchmark cases, the relative depth reduction increases with respect to qubit size, indicating that the optimization strategy scales well with circuit size of these circuits.

To further validate the robustness of the optimization strategy, we performed a large-scale benchmark experiment consisting of 1000 synthetic, yet structured quantum circuits containing distributed CNOT operations. These circuits were generated using ketGPT, a tool designed to generate circuits that mimic the structural patterns and characteristics typically found in quantum circuits. 

The motivation behind this experiment is to evaluate the compiler on a much larger and more diverse set of circuits than the initial benchmark suite. While the earlier experiments demonstrate that the compiler can reduce circuit depth for specific benchmark circuits, this larger evaluation aims to verify that the optimization strategy behaves reliably across many circuit structures and consistently avoids transformations that could increase circuit depth.

\begin{figure}
    \centering
    \includegraphics[width=1\linewidth]{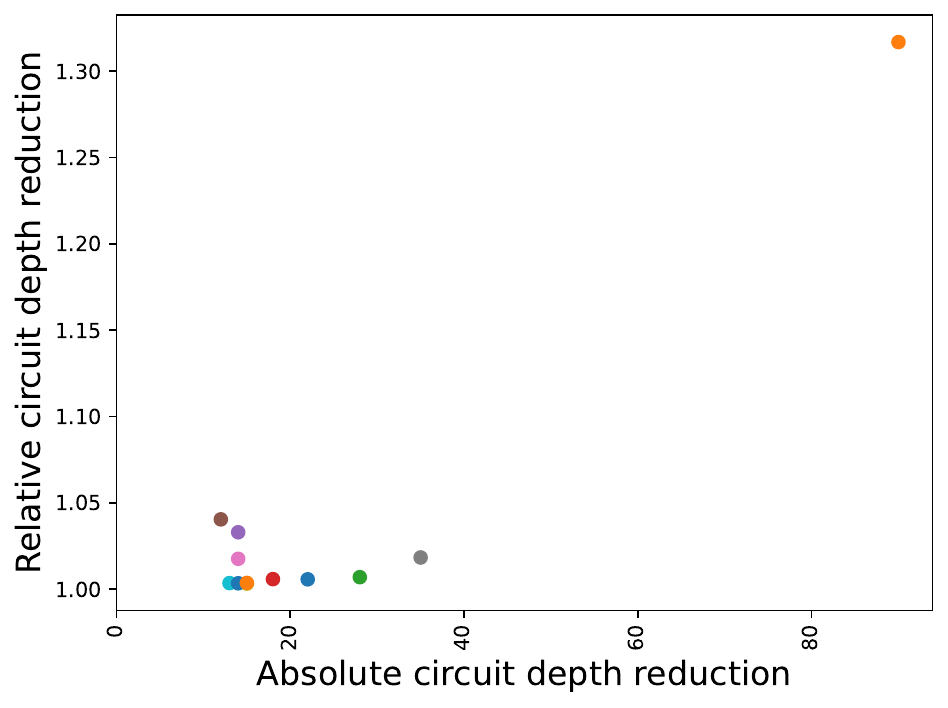}
    \caption{Absolute and relative circuit depth improvements obtained by the compiler across 1000 synthetic, yet structured benchmark circuits generated with ketGPT. Each dot represents a benchmark circuit that has achieved an improvement, circuits that remained equal are left out of the plot for readability. The results show that the compiler identifies depth reduction opportunities while never increasing circuit depth.}
    \label{fig:ketgpt_absolute_relative}
\end{figure}

The results of this experiment are shown in Figure \ref{fig:ketgpt_absolute_relative}, which shows the absolute circuit depth improvement against the relative improvement achieved by the compiler.

Two important observations can be made from these results. First, even across this large and diverse set of circuits, the compiler never increases the circuit depth of any benchmark. This strengthens the confidence that the optimization strategy is conservative and only applies transformations when they are guaranteed to be beneficial. Second, the compiler still identifies opportunities for improvement, achieving depth reductions across a portion of the generated circuits.


As mentioned before, to guarantee that the compiler never degrades circuit performance, the optimization strategy is intentionally designed to be conservative. Transformations are only applied when they are guaranteed not to increase the circuit depth. While this ensures that the compiler never produces a circuit that is worse than the original decomposition, it also means that some potential optimization opportunities are intentionally not exploited.

\begin{figure}
    \centering
    \includegraphics[width=1\linewidth]{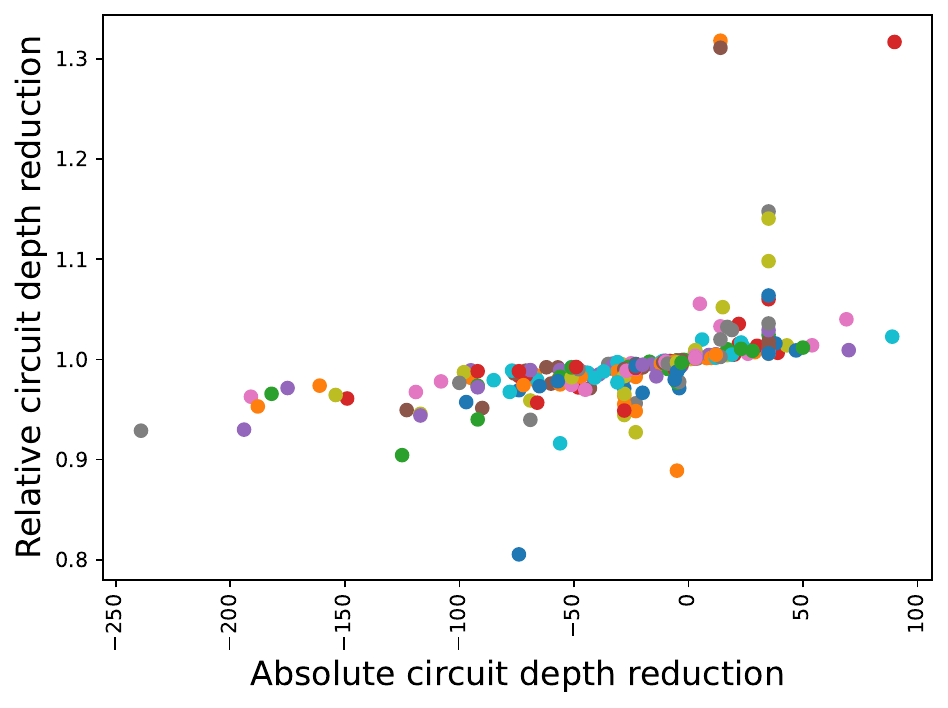}
    \caption{Absolute and relative circuit depth results obtained by the compiler across 1000 synthetic, yet structured benchmark circuits generated with ketGPT, while not running a conservative approach. Each dot represents a benchmark circuit that has achieved a change in circuit depth, circuits that remained equal are left out of the plot for readability. The results show that the compiler, when not run conservatively, identifies depth reduction opportunities that can either result in larger depth reductions compared to Fig \ref{fig:ketgpt_absolute_relative}, but can also result in depth increases.}
    \label{fig:non_conservative_results}
\end{figure}

To illustrate this trade-off, we performed an additional experiment in which the conservative constraints of the compiler were relaxed. Figure \ref{fig:non_conservative_results} shows the resulting absolute and relative circuit depth changes. As expected, relaxing these constraints allows the compiler to identify additional opportunities for depth reduction, leading to larger improvements in some cases. However, this also introduces cases where the circuit depth increases.

These results highlight the rationale behind the conservative design of the current compiler: it guarantees safe optimizations while still achieving meaningful improvements in many circuits. At the same time, the experiment suggests that more advanced compiler implementations could potentially achieve even larger depth reductions in the future.

\section{Conclusion}\label{sec:conclusion}

This work introduced a compiler and theory framework that performs depth-aware logical-to-physical decomposition for quantum circuits, with a particular focus on distributed CNOT operations. By combining instruction rescheduling, and hardware-aware decomposition, the compiler uncovers parallelism in logical CNOT gates that either share control or target qubits. The key insight driving this approach is that many logical CNOT sequences sharing a control or target qubit, can be executed in parallel at the physical level using shared entanglement resources.


The compiler uses a multi-pass bucket merging strategy that ensures the opportunities for performing decompositions are detected. Because the compiler only applies transformations that preserve or reduce circuit depth, it guarantees that the optimized circuit is never worse than a naive decomposition.

We have evaluated our compiler by running it over a set of benchmarks. These benchmark results confirm the expected behavior: circuits containing inherently sequential CNOT structures see substantial depth reductions, while circuits that are already maximally parallel remain unchanged.

These findings highlight the importance of integrating scheduling and decomposition into an optimization process. Treating these steps independently misses opportunities for parallel execution that are essential for efficient distributed quantum computing. Our approach directly targets the depth overhead introduced by distributed operations, making it particularly well-suited for NISQ-era devices where every layer of depth contributes to noise.



In conclusion, our compiler provides a foundation for further advances in the optimization of distributed quantum circuits. The current design intentionally adopts a conservative optimization strategy, ensuring that compiler transformations never increase circuit depth. This approach guarantees safe optimizations, but it also leaves some potential improvements unexplored. Experiments with a relaxed version of the compiler indicate that additional depth reductions are achievable, which creates opportunities for further compiler optimization strategies.

Future work may also explore architectures with more complex connectivity patterns, such as systems that do not offer all-to-all communication and therefore require additional routing or entanglement-distribution strategies. Other promising directions include supporting richer multi-qubit entanglement structures or scenarios in which multiple logical qubits reside on the same physical node. As distributed architectures continue to evolve, compilers capable of exploiting shared entanglement and structural parallelism will play a crucial role in enabling scalable, high-fidelity quantum computation.

\section{Acknowledgments}
We gratefully acknowledge support from the joint research program “Modular quantum computers” by Fujitsu Limited and Delft University of Technology, co-funded by the Netherlands Enterprise Agency under project number PPS2007.

\bibliographystyle{IEEEtran}
\bibliography{report.bib}

@IEEEtranBSTCTL{IEEEexample:BSTcontrol,
  CTLuse_forced_etal       = "yes",
  CTLmax_names_forced_etal = "2",
  CTLnames_show_etal       = "1" 
}

@article{Arute2019QuantumSupremacy,
  title={Quantum supremacy using a programmable superconducting processor},
  author={Arute, Frank and Arya, Kunal and Babbush, Ryan and others},
  journal={Nature},
  volume={574},
  number={7779},
  pages={505--510},
  year={2019},
  publisher={Nature Publishing Group}
}

@article{Zhong2020BosonSampling,
  title={Quantum computational advantage using photons},
  author={Zhong, Han-Sen and Wang, Hui and Deng, Yu-Hao and others},
  journal={Science},
  volume={370},
  number={6523},
  pages={1460--1463},
  year={2020},
  publisher={American Association for the Advancement of Science}
}

@article{Monroe2014LargeScale,
  title={Large-scale modular quantum-computer architecture with atomic memory and photonic interconnects},
  author={Monroe, Christopher and Kim, Jungsang},
  journal={Physical Review A},
  volume={89},
  number={2},
  pages={022317},
  year={2014},
  publisher={APS}
}

@article{Nickerson2014DQC,
  title={Freely scalable quantum technologies using cells of 5-to-50 qubits with very lossy and noisy photonic links},
  author={Nickerson, Naomi H. and Fitzsimons, Joseph F. and Benjamin, Simon C.},
  journal={Nature Communications},
  volume={4},
  pages={1756},
  year={2013},
  publisher={Nature Publishing Group}
}

@article{Wehner2018QuantumInternet,
  title={Quantum internet: A vision for the road ahead},
  author={Wehner, Stephanie and Elkouss, David and Hanson, Ronald},
  journal={Science},
  volume={362},
  number={6412},
  pages={eaam9288},
  year={2018},
  publisher={American Association for the Advancement of Science}
}

@article{Preskill2018NISQ,
  title={Quantum Computing in the NISQ era and beyond},
  author={Preskill, John},
  journal={Quantum},
  volume={2},
  pages={79},
  year={2018}
}

@article{Kandala2017HardwareEfficient,
  title={Hardware-efficient variational quantum eigensolver for small molecules and quantum magnets},
  author={Kandala, Abhinav and Mezzacapo, Antonio and Temme, Kristan and others},
  journal={Nature},
  volume={549},
  number={7671},
  pages={242--246},
  year={2017},
  publisher={Nature Publishing Group}
}

@inproceedings{Murali2019NoiseAware,
  title={Noise-adaptive compiler mappings for noisy intermediate-scale quantum computers},
  author={Murali, Prakash and Baker, Jonathan M. and Javadi-Abhari, Ali and others},
  booktitle={Proceedings of the 24th International Conference on Architectural Support for Programming Languages and Operating Systems},
  pages={1015--1029},
  year={2019}
}

@article{Cowtan2019QubitMapping,
  title={Qubit allocation and routing via max-flow/min-cut},
  author={Cowtan, Alexander and Dilkes, Sam and Duncan, Ross and others},
  journal={Quantum Science and Technology},
  volume={5},
  number={4},
  pages={044015},
  year={2020}
}

@article{Hensen2015LoopholeFree,
  title={Loophole-free Bell inequality violation using electron spins separated by 1.3 kilometres},
  author={Hensen, Bas and Bernien, Hannes and Dréau, Anaïs E. and others},
  journal={Nature},
  volume={526},
  number={7575},
  pages={682--686},
  year={2015},
  publisher={Nature Publishing Group}
}

@article{Pompili2021Multinode,
  title={Realization of a multinode quantum network of remote solid-state qubits},
  author={Pompili, Matteo and Hermans, Sarah L. N. and Baier, Sebastian and others},
  journal={Science},
  volume={372},
  number={6539},
  pages={259--264},
  year={2021},
  publisher={American Association for the Advancement of Science}
}

@article{Chamberland2022Topological,
  title={Building a fault-tolerant quantum computer using concatenated cat codes},
  author={Chamberland, Christopher and Noh, Kyungjoo and Arrangoiz-Arriola, Patricio and others},
  journal={PRX Quantum},
  volume={3},
  number={1},
  pages={010329},
  year={2022},
  publisher={APS}
}

@inproceedings{Amy2013,
  title={A meet-in-the-middle algorithm for fast synthesis of depth-optimal quantum circuits},
  author={Amy, Matthew and Maslov, Dmitri and Mosca, Michele},
  booktitle={IEEE Transactions on Computer-Aided Design of Integrated Circuits and Systems},
  volume={32},
  number={6},
  pages={818--830},
  year={2013},
  publisher={IEEE}
}

@article{Smith2019Qiskit,
  title={An open-source framework for quantum computing},
  author={Aleksandrowicz, Gadi and Alexander, Thomas and Barkoutsos, Panagiotis and others},
  journal={Qiskit: IBM Research},
  year={2019},
  note={https://qiskit.org}
}

@article{Sivarajah2020Tket,
  title={t|ket⟩: A retargetable compiler for NISQ devices},
  author={Sivarajah, Shouvanik and Dilkes, Sam and Cowtan, Alexander and others},
  journal={Quantum Science and Technology},
  volume={6},
  number={1},
  pages={014003},
  year={2020}
}

@article{Li2020QubitMapping,
  title={Qubit mapping via multi-level intermediate representation},
  author={Li, Gushu and Ding, Yongshan and Xie, Yuan},
  journal={Proceedings of the 2020 IEEE/ACM International Symposium on Microarchitecture},
  pages={69--82},
  year={2020}
}

@article{Zhou2020Commutation,
  title={Quantum circuit optimization using commutativity and gate identities},
  author={Zhou, Li and Wang, Jian and Ying, Mingsheng},
  journal={Physical Review A},
  volume={102},
  number={6},
  pages={062412},
  year={2020}
}

@inproceedings{Chakrabarti2021SABRE,
  title={SABRE: A depth-optimized quantum mapping tool for NISQ systems},
  author={Chakrabarti, Shashank and Yoder, Theodore J. and others},
  booktitle={Proceedings of the 2021 Design Automation Conference},
  pages={1--6},
  year={2021}
}

@article{Tan2023DepthOptimized,
  title={Depth-optimized quantum circuit synthesis via hierarchical rewriting},
  author={Tan, Bochen and Lin, Chia-Chun and others},
  journal={npj Quantum Information},
  volume={9},
  number={1},
  pages={1--12},
  year={2023}
}

@article{VanMeter2016,
  title={Quantum networking and distributed quantum computing},
  author={Van Meter, Rodney},
  journal={IEEE Network},
  volume={30},
  number={5},
  pages={26--32},
  year={2016}
}

@article{Maslov2018,
  title={Basic circuit compilation techniques for an ion-trap quantum machine},
  author={Maslov, Dmitri},
  journal={New Journal of Physics},
  volume={19},
  number={2},
  pages={023035},
  year={2017},
  publisher={IOP Publishing}
}

@article{Nash2020,
  title={Quantum circuit optimizations for variational algorithms},
  author={Nash, Bryan and Gheorghiu, Vlad and Mosca, Michele},
  journal={Quantum Science and Technology},
  volume={5},
  number={2},
  pages={025010},
  year={2020}
}

@inproceedings{Zulehner2018,
  title={An efficient methodology for mapping quantum circuits to the IBM QX architectures},
  author={Zulehner, Alwin and Paler, Alexandru and Wille, Robert},
  booktitle={Proceedings of the 2018 Design, Automation \& Test in Europe Conference},
  pages={1135--1138},
  year={2018}
}

@article{Childs2019,
  title={Circuit transformations for quantum architectures},
  author={Childs, Andrew M. and Schoute, Eddie and Unsal, Cem},
  journal={arXiv preprint arXiv:1902.09102},
  year={2019}
}

@article{Cirac1999,
  title={Distributed quantum computation over noisy channels},
  author={Cirac, J. Ignacio and Zoller, Peter and Kimble, H. Jeff and Mabuchi, Hideo},
  journal={Physical Review Letters},
  volume={78},
  number={16},
  pages={3221--3224},
  year={1997},
  publisher={APS}
}

@article{Caleffi2018,
  title={End-to-end entanglement rate: Toward a quantum route metric},
  author={Caleffi, Marcello},
  journal={IEEE Transactions on Quantum Engineering},
  volume={1},
  pages={1--14},
  year={2020}
}

@inproceedings{Nam2018,
  title={Automated optimization of large quantum circuits with continuous parameters},
  author={Nam, Yunseong and Ross, Neil J. and Su, Yuan and others},
  booktitle={Proceedings of the 2018 International Symposium on Computer Architecture},
  pages={194--205},
  year={2018}
}

@article{Younis2022,
  title={QFAST: Quantum circuit synthesis using a fast search},
  author={Younis, Ahmed and Shende, Vivek and others},
  journal={Quantum},
  volume={6},
  pages={761},
  year={2022}
}

@article{Bertoni2023,
  title={Deep optimization of quantum circuits with reinforcement learning},
  author={Bertoni, Claudio and Caro, Matteo C. and others},
  journal={npj Quantum Information},
  volume={9},
  number={1},
  pages={1--10},
  year={2023}
}

@article{Nam2020,
  title={Ground-state energy estimation of the water molecule on a trapped-ion quantum computer},
  author={Nam, Yunseong and Chen, Jwo-Sy and Pisenti, Neil and others},
  journal={npj Quantum Information},
  volume={6},
  number={1},
  pages={1--6},
  year={2020}
}

@inproceedings{ketGPT,
author = {Apak, Boran and Bandic, Medina and Sarkar, Aritra and Feld, Sebastian},
title = {KetGPT – Dataset Augmentation of Quantum Circuits Using Transformers},
year = {2024},
isbn = {978-3-031-63777-3},
publisher = {Springer-Verlag},
address = {Berlin, Heidelberg},
url = {https://doi.org/10.1007/978-3-031-63778-0_17},
doi = {10.1007/978-3-031-63778-0_17},
booktitle = {Computational Science – ICCS 2024: 24th International Conference, Malaga, Spain, July 2–4, 2024, Proceedings, Part VI},
pages = {235–251},
numpages = {17},
location = {Malaga, Spain}
}

@article{quetschlich2023mqtbench,
  title        = {{MQT Bench}: Benchmarking Software and Design Automation Tools for Quantum Computing},
  shorttitle   = {MQT Bench},
  author       = {Quetschlich, Nils and Burgholzer, Lukas and Wille, Robert},
  journal      = {Quantum},
  volume       = {7},
  pages        = {1062},
  year         = {2023},
  doi          = {10.22331/q-2023-07-20-1062},
  eprint       = {2204.13719},
  eprinttype   = {arxiv},
  note         = {{MQT Bench} is available at \url{https://www.cda.cit.tum.de/mqtbench/}}
}


\end{document}